\documentclass[%
 aip,
 amsmath,amssymb,
 reprint,%
groupedaddress
]{revtex4-1}

\usepackage{graphicx}
\usepackage{dcolumn}
\usepackage{bm}

\usepackage[utf8]{inputenc}
\usepackage[T1]{fontenc}
\usepackage{mathptmx}
\usepackage{etoolbox}
\usepackage{amsthm}
\usepackage{xcolor}
\makeatletter
\def\@email#1#2{%
 \endgroup
 \patchcmd{\titleblock@produce}
  {\frontmatter@RRAPformat}
  {\frontmatter@RRAPformat{\produce@RRAP{*#1\href{mailto:#2}{#2}}}\frontmatter@RRAPformat}
  {}{}
}%
\makeatother

\newtheorem*{theorem*}{Framework}

\begin{document}

\preprint{AIP/123-QED}
\title{Perfect Superconducting Diode and Supercurrent Range Controller}
\author{Cliff Sun}
\author{Ziqi Zhao}
\author{Alexey Bezryadin}%
\affiliation{University of Illinois at Urbana-Champaign, Department of Physics}

\begin{abstract}
Diodes have a nonreciprocal voltage versus current relationship, produced by breaking the space and time reversal symmetry. However, developing high-end superconducting computers requires a superconducting analogue of the traditional semiconductor diode. Such a superconducting diode exhibits non-reciprocity, or a high asymmetry in its critical currents. We present a model of a perfect superconducting diode based on a superconducting quantum interference device made with multiple superconducting nanowires. The diode predicted by our model has a large positive critical current, while the negative critical current can be exactly zero. This 100\% diode efficiency ($\eta = 1$) remains stable against small changes of the magnetic field. Another important result is that under certain and quite broad conditions such devices can act as supercurrent range controllers. In such device a supercurrent can flow with zero voltage applied, but only if the supercurrent is contained in some narrow, adjustable range, which excludes zero current.  

\end{abstract}
\maketitle


\section{Introduction}

Superconducting electronics offer a significant energy advantage over that of their traditional semiconductor counterpart due the absence of resistivity \cite{tinkham_book, Bezryadin_book}. This drastically improves energy scalability of superconducting computing resources.
 
Superconducting diodes are useful for quantum and cryogenic electronics with their ability to allow supercurrent to flow preferentially in one direction \cite{PhysRevLett.129.267702, PhysRevB.98.075430, PhysRevB.98.054510, baumgartner-2021}. In quantum computing, superconducting diodes are crucial for efficient power delivery in quantum processors, ensuring low-noise operation at ultra-cold temperatures \cite{ingla-aynes-2025, ideue2020one}. In cryogenic electronics, superconducting diodes enable rectification circuits that convert AC to DC in cryogenic temperatures which improves energy efficiency \cite{tuhin-2025}. Unique properties of superconducting diodes make them ideal for ultra-sensitive detection systems and sensors used in medical imaging \cite{ando-2020, PhysRevLett.131.027001}. Superconducting diodes might play a role in advanced superconducting circuits and enable the development of the next-generation superconducting logic devices and memory elements \cite{ma-2025, alam-2021, eduard_memory, song-2023}. Note that an intrinsic perfect superconducting diode (PSD) has been discussed theoretically \cite{chakraborty-2024}. Additionally, Josephson Junctions have been used, but are limited due to their large size \cite{9827535, herr-2023}. Therefore, new approaches are necessary\cite{golod-2023}. We propose an alternative approach to construct PSD using one-dimensional superconducting nanowires which doesn't require any exotic superconductor and can be made out of ordinary superconducting metals such as niobium or aluminum. 

We present a simple yet efficient model of a superconducting quantum interference device (SQUID) made with a few one-dimensional nanowires acting as parallel weak links. One key difference with respect to the ordinary SQUID is that our nanowires are assumed to have a linear current-phase relationship, as established previously\cite{murphy-2017, murphy,eduard_memory, Bezryadin_book}. The phase difference $\phi$ is limited by the critical phase $\phi_c$, proportional to the wire length. The linear CPR means that the supercurrent, $I$, is proportional to the phase gradient of the condensate wave function (similar to free quantum particle). The phase gradient, on the other hand, is proportional to the phase difference $\phi$ between the ends of the nanowire. The linear CPR enables the manifestation of multiple highly stable vorticity states. Our model\cite{model1,model2} allows us to calculate vorticity stability regions (VSR), which are the regions in the current-magnetic (I-B) field plane within which the total supercurrent is within the negative and positive critical current. We discover that some of the vorticity states generate VSRs that act as an supercurrent range controller (SRC) or as a perfect superconducting diode (PSD). In an SRC, the supercurrent ($I$) is confined between two positive values, i.e. $0 < I_{c,-} < I < I_{c,+}$. In a PSD, the supercurrent is confined as $0 = I_{c,-} < I < I_{c,+}$, which corresponds to the perfect efficiency $\eta=(I_{c,+}+I_{c,-})/(I_{c,+}-I_{c,-})=1$. Note that since in the SRC $I_{c,-}>0$, therefore formally speaking $\eta>1$.

We will show that there exists two forms of the PSD: point-wise PSD and b-invariant PSD. The former is a device that acts like a PSD for specifically applied magnetic fields. A b-invariant PSD is a device that acts like a PSD regardless of the applied magnetic field (in a certain range).
Both the SRC and the PSD are achieved by breaking space and time symmetry of the SQUID. However, as we will show later, a b-invariant PSD can only form in devices with 3 or more superconducting nanowires. An SRC requires a minimum of two nanowires. We also show that a PSD can be transformed into SRC by introducing differences (disorder) between the nanowires. 

The results presented can apply to other types of interferometers in which matter waves are involved. For quantum gases for example, the probability current is linearly proportional to the phase gradient. Thus our results can pertain to atomic interferometers \cite{berman1997atom}. 

\section{Linear CPR model for a multiple wire SQUID}

We consider a parallel array of superconducting nanowires connected to two macroscopic thin-film superconducting electrodes. To model this device, we assume that all the superconducting nanowires obey a linear\cite{tinkham_book} current-phase-relationship (CPR) as given by: 

\begin{equation}
    I_i = I_{c,i}\frac{\phi_i}{\phi_{c,i}}
    \label{e1}
\end{equation}

where $i$ is the wire number, $\phi_i$ is the phase difference across the wire $i$, and $\phi_{c, i}$ and and $I_{c,i}$ represent the critical phase and the critical current of the wire $i$. In other words, if the current in the wire $i$ equals its critical current, $I_{c,i}$, then the phase across the wire is defined as the critical phase, $\phi_{c, i}$, of the wire. The phase across the wire (or the phase bias of the wire) is defined as the phase difference of the superconducting condensate complex wave function, taken between the ends of the wire. The ends of the wire are the points where the wire connects to macroscopic superconducting electrodes.

\section{Switching current criterion}
In this section we elucidate at which point our multi-wire superconducting quantum interference device (MW-SQUID) switches to the normal state. To confirm our conclusions we add experimental results, which illustrates that our choice for the critical current criterion for the entire device is reasonable. 

The starting point for our model is the notion that each wire has a critical current $I_{c,i}$, such that if the bias current in the wire, $I_i$, is larger than $I_{c,i}$ then the wire switches to the resistive, normal state, i.e., it becomes non-superconducting. The key criterion of the model is that the critical current of the entire device (MW-SQUID) is reached if the current in any of the wires reaches its critical value for that wire. Suppose the bias current of the entire device is $I$, and the critical current of the device is $I_{c}(v_k)$, defined for a given fixed distribution of vortices $v_k$ in the cells between the wires. Here $k$ (integer) is the cell number. With these notations, the critical current criterion of our model is $I=I_c(v_k)$ if  $I_i=I_{c,i}$ for any wire number $i$. Since the vorticity, $v_k$, can accept different values, one expects that the critical current will be a multi-valued function of the magnetic field. This is confirmed experimentally as is shown below.  

Here we discuss in detail the critical current criterion, stating that $I=I_c(v_k)$ if $I_i=I_{c,i}$. The key idea is that the critical current function, $I_c(v_k)$, for a given vorticity state represents the boundary of the corresponding vorticity stability region (VSR). To understand this, consider the case when the current is ramped up from zero. At some time the current in one wires becomes equal to the critical current of this wire $i$, $I_i=I_{c,i}$. Then the wire cannot remain superconducting. Then one of two scenarios might happen. (1) If there is some other vorticity state in which the currents in all wires are subcritical (for the given total bias current $I$) then there will be a phase slip in the wire which case the numbers of vortices trapped between the wires to change in such a way as to reduce the current in the wire $i$. As a result of such phase slip some heat is generated and the entire device typically also switches to a non-superconducting regime, since its temperature is increased. (2) If the bias current is low then the heat generated might be insufficient to switch the entire device. Then the event of vortex redistribution is classified as a hidden phase slip. Such hidden phase slips can be observed by using a special technique reported in Murphy et al.\cite{murphy-2017}. In both cases, the current vorticity state has been changed or destroyed. This means that the boundaries of the VSR is the critical current for that specific vorticity.

\begin{figure}
    \centering
    \includegraphics[width=\linewidth]{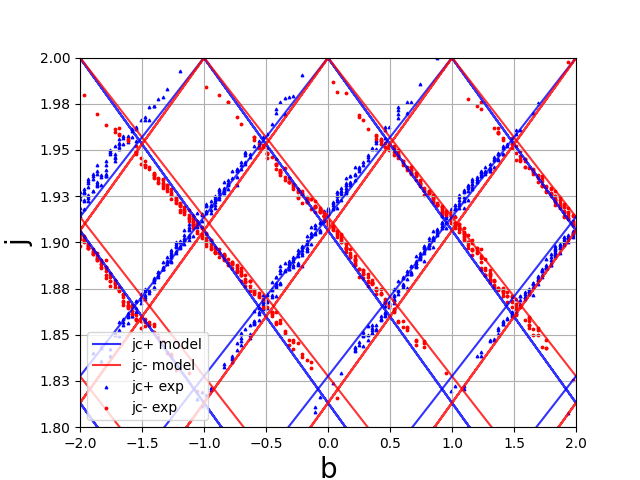}
    \caption{\textbf{2-SQUID experimental data with model fits.} The model parameter fits can be found in Table~\ref{tab:parameters}. This experimental data and our model fits validates our vorticity switching assumption, where if one nanowire reaches its critical current, then the vorticity of the device will switch due to a phase slip. }
    \label{fig:model_fit}
\end{figure}


It is clear that a given fixed vorticity distribution is stable only as long as all wires remain subcritical ($I_i<I_{c,i}(v_k)$). As soon as one of the wire reaches its critical point ($I_i=I_{c,i}(v_k)$) the vorticity state must change. Such reshuffling of vortices might cause the device to switch to the normal state with some probability. To compare with experimental results, we calculate critical currents for many vorticity states and thus obtain a multi-valued function $I_c(v_k, B)$. Then some branches of this function can be observed to correspond to the switching events of the entire device. However, some branches might not manifest themselves under typical experimental conditions.

\begin{figure}
    \centering
    \includegraphics[width=\linewidth]{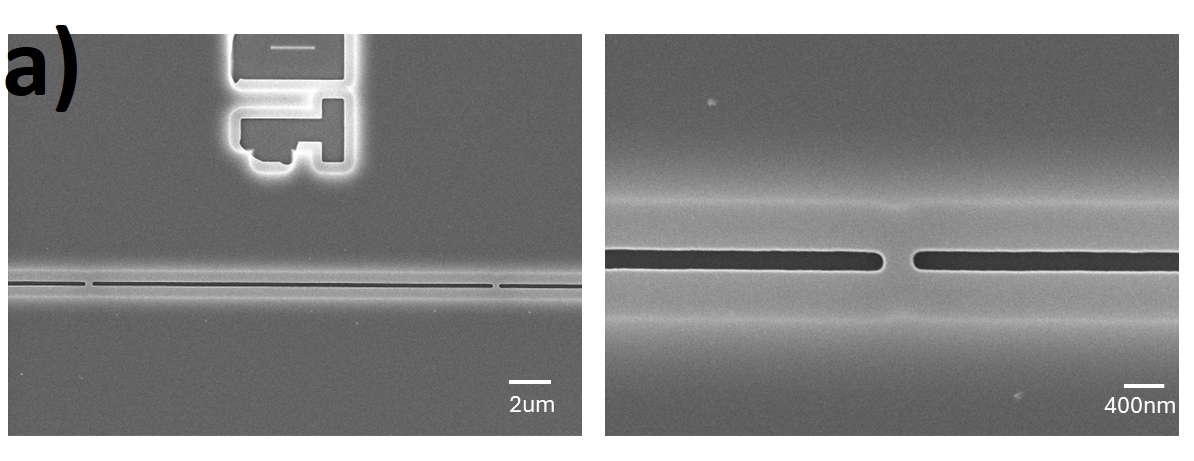}
    \includegraphics[width=\linewidth]{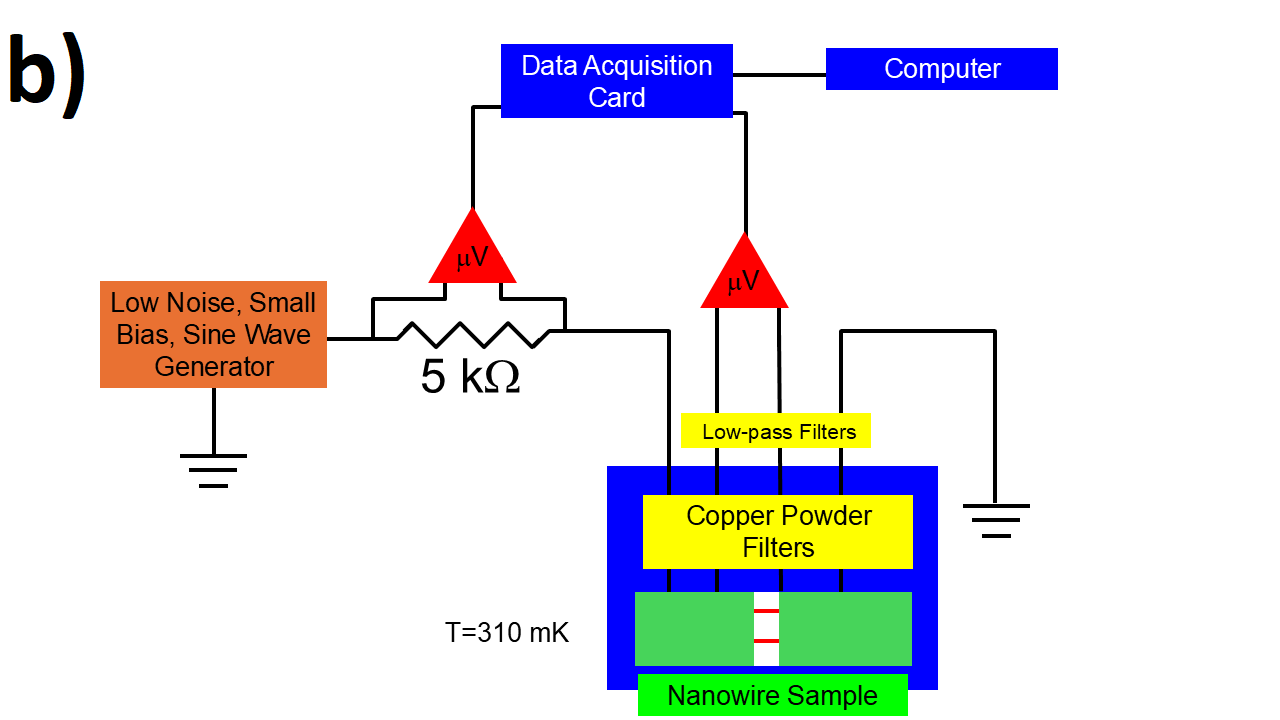}
    \caption{\textbf{(a)} \textbf{SEM image of the superconducting 2-nanowire sample} The left image is an overview image of the nanowires sample. The right image is a zoomed-in image of one of the fabricated superconducting nanowires.  \textbf{(b)} \textbf{Schematic of the four-probe measurement circuit.} A voltage source is put in series with a high-value standard resistor $R_{st}$ (shown 5k$\Omega$ here) compared to the sample resistance. This set-up produces a small (relative to zero) current bias to the nanowire sample. An NI-DAQ card equipped with an analog-to-digital converter simultaneously measures the sample voltage, V, and the voltage across the standard resistor, $V_{st}$. These measurements are recorded in real time as the bias current is swept up and down. The bias current is then calculated using Ohm’s law, $I = V_{st}/R_{st}$. The resulting V-I data points are plotted on the computer screen. Both V and $V_{st}$ signals are amplified by preamplifiers (red triangles) before entering the NI-DAQ card input channels. The critical current is defined as the current at which a sharp jump is observed from $V=0$ to $V>0$.}
    \label{fig:setup.png}
\end{figure}

We validated this switching assumption by fabricating a simple 2-SQUID, i.e., a SQUID with two superconducting nanobridges, parallel to each other and connecting larger electrodes. We measure the critical current of the device versus perpendicular magnetic field, and we performing fits using our model. The 2-SQUID sample was fabricated on Si substrate covered with oxide $SiO_{2}$ and a low stress $SiN$ on top. A trench with two bridges was fabricated using standard electron beam lithography and reactive ion etching. A hydrofluoric acid etch selectively removed the underlying $SiO_{2}$, creating an undercut that fully suspends the $SiN$ nanobridges. Photolithography was subsequently employed to define electrode patterns, followed by $Nb$ deposition via sputtering and a lift-off process to complete the device structures. A zoomed out SEM image of the sample can be found in Fig.~\ref{fig:setup.png}a. A zoomed in image on one of the bridges is shown in Fig.~\ref{fig:setup.png}b. Our measurement set-up is illustrated in Fig.~\ref{fig:setup.png}c. It contains a controllable voltage source and a resistor connected in series with the sample to realized a current biasing experiment. Voltage is measured on the series resistor (to calculate the current by Ohm's law) and on the sample. The voltage-current curves are then plotted in LabView. They normally show a voltage jump at some current, which is then taken as the critical current of the entire device. 

It is important to note that we have normalized the critical currents such that $j_{c,i} = I_{c,i}/\langle I_{c,i}\rangle $ where $\langle I_{c,i} \rangle$ is the average of all the critical currents, $\langle I_{c,i} \rangle=\sum_i I_{c,i}/n$. This is so $\sum_i j_{c,i} = n\langle I_{c,i} \rangle / \langle I_{c,i} \rangle = n$ where $n$ is the number of wires in the SQUID. In the example considered in this section $n=2$. Therefore, the maximum normalized critical current of the entire device equals $2$ .  

\begin{table}[h] 
\centering
\begin{tabular}{|c|c|c|c|c|}
\hline
$\phi_{c,1}$ (radians) & $\phi_{c,2}$ (radians) & $j_{c,1}$ & $j_{c,2}$ & $\Delta B$ (Gauss) \\ \hline
70 & 70 & 1.05 & 0.97 & 0.027 \\ \hline
\end{tabular}
\caption{Model Fitting Parameters}
\label{tab:parameters}
\end{table}

The measured critical current curves are shown in Fig.~\ref{fig:model_fit}. The positive critical current is shown by blue triangles and the negative critical current is shown by red dots. The positive and negative critical currents were normalized by dividing by half of the maximum critical current($I_{max} = 1.18 mA)$, corresponding to the average value of all measured critical currents. The experimental magnetic field was then referenced to zero field by exploiting the intrinsic symmetry of the $I_c$–$B$ pattern and shifting the $x$-axis such that $B=0$ satisfied this symmetry. Finally, the field axis was scaled by $\delta B$ to align the experimental data with the model.The function of the critical current plotted versus magnetic field appears multi-valued. This fact reflects the expectation that different vorticity states have different critical currents. The corresponding fits produced by our model are shown by blue and red lines (Fig.~\ref{fig:model_fit}). Note, the experimental data and the model's total critical current were normalized. The data fit demonstrates strong evidence that our model assumption, the criterion for the switching of the device, is a good description of such nanowire SQUIDs. 

In Fig.~\ref{fig:model_fit}, the experimental data branches corresponds to the right sides of the diamond VSRs. Each branch corresponds to a unique vorticity state. Our switching assumption states that if the critical current of the VSR is reached, then the device will switch its vorticity if and only if there is a vorticity state for which the current in the critical wire is less than its critical current (for a given magnetic field). On the contrary, if $I_c(v_{k}, B) > I_{c}(v^{'}_{k}, B)$, for any other vorticity distribution $v^{'}_{k}$ then the device won't switch its vorticity state, but rather turn normal. But if $I_c(v_{k},B) < I_{c}(v^{'}_{k+1},B)$, then the device switches to $v^{'}_{k}$. With a nonzero probability such vorticity change cause enough dissipation and switch the device to the normal state, especially if the current is near the critical current. The fact that we observe a multi-valued function (after many repetitive measurements of the critical current gives us evidence that our interpretation that a single phase slip can switch the device is correct. Moreover, the observed agreement between the fits and the data confirms our model.

\section{Effect of the magnetic field and vortices on the critical current}

If a magnetic field is applied perpendicular to the device, the Meissner currents produce additional phase shifts between wires. Each vortex positioned between the wires reduces the phase shift between the corresponding wires by $2\pi$. Thus, the Meissner phase correlation (MPC) equation for wires $i$ and $j$ is\cite{hopkins_SQUID,pekker-2005}:

\begin{equation}
    \phi_{j}=\phi_i+2\pi \left(\frac{B}{\Delta B}\right)\left(\frac{X_j - X_i}{X_n - X_1}\right)-2\pi v_{i,j}
    \label{e2}
\end{equation}

Here, $X_i$ is the x-coordinate of the wire $i$ and $B$ is the magnetic field. The nanowires are numbered from left to right, from $1$ to $n$. If $j>i$ then $v_{i,j}$ is the number of vortices between wires $i$ and $j$, which is an integer. We define $v_{i,j} = -v_{j,i }$. For example, if there are three wires and there is one vortex in the first cell and two vortices in the second cell then $v_{1,2}=1$, $v_{2,3}=2$ and so $v_{1,3}=1+2=3$, and $v_{2,1}=-1$, and $v_{2,1}+v_{2,3}=-1+2=1$. 

Next, we normalize the parameters $I_{c,i}$, $B$ and $X_i$. First note that the period of the Little-Parks oscillations\cite{little_parks, gurt-yo,Belkin2011-nv} for a SQUID with just two wires (2-SQUID)\cite{hopkins_SQUID, pekker-2005} (illustrated in Fig.\ref{fig:Sample design}) is $\Delta B=(\pi^2/8G)(\phi_0/[W(X_2-X_1)])$ where $G$ is the Catalan number. We define a constant $\Delta B_1 =(\pi^2/8G)(\phi_0/W)$, where $W$ is the width of the electrodes of the SQUID. Then the period of the 2-SQUID is $\Delta B=\Delta B_1/(X_2-X_1)$. Suppose for now that there are no vortices. Then, the MPC equation is $\phi_2=\phi_1+2\pi B/\Delta B-2\pi v_{i,j}$, by the definition of the period $\Delta B$. It can also be written as $\phi_2=\phi_1+2\pi (X_2-X_1)B/\Delta B_1-2\pi v_{i,j}$. If there are $n$ parallel wires (n-SQUID) then the MPS equation is generalized as $\phi_i=\phi_1+2\pi (X_i-X_1)B/\Delta B_1-2\pi v_{i,j}$, where $X_i$ is the position of the wire number $i$.

For a SQUID with $n$ wires (n-SQUID) the normalized magnetic field is defined as $b=B/[\Delta B_1/(X_n-X_1)]$ and the normalized coordinates of the wires are defined as $x_i=(X_i-X_1)/(X_n-X_1)$. Therefore $x_ib=\{(X_i-X_1)/(X_n-X_1)\} \times \{B[(X_n-X_1)/\Delta B_1]\}=(X_i-X_1)B/\Delta B_1$. Then the MPC equation becomes $\phi_i=\phi_1+2\pi x_ib-2\pi v_{i,j}$. Similarly, for wires $i$ and $i+1$, the MPC equation becomes $\phi_{i+1}=\phi_i+2\pi b(x_{i+1} - x_i)-2\pi v_{i,j}$. Each vortex present between wires $i$ and $i+1$ reduced the phase shift by $2\pi$. Finally, the MPC equation for the wires $i$ and $j$ is 

\begin{equation}
    \phi_{j}=\phi_i+2\pi b (x_{j}-x_i)-2\pi v_{i,j}
    \label{e3}
\end{equation}

It is important to note that the definition of $x_i$ is such that $x_1 = 0$ always and $x_n = 1$ where $n$ is the order number of the right-most wire in the device.
The normalized supercurrent in wire $i$ is:

\begin{equation}
    j_{i} = j_{c,i}\frac{\phi_i}{\phi_{c,i}}
    \label{e3b}
\end{equation}

Where $j_i = I_i/\langle I_{c,i} \rangle$ is the normalized supercurrent in wire $i$ and $\langle I_{c,i} \rangle$ is the average of the critical currents of the wires. Also, $j_{c,i}$ is the normalized critical current of wire $i$ defined as $j_{c,i} = I_{c,i}/\langle I_{c,i} \rangle$.

Using Eqs.\eqref{e3},\eqref{e3b} we generate vorticity stability regions (VSR), each defined as region on the $I-B$ plane where for a fixed vorticity, the current in each wire is lower than its corresponding critical current. The boundary of the VSR is the critical current of the device as a function of the magnetic field. If either the total current, $I$, or the magnetic field, $B$ is changed beyond the VSR boundaries, then the device switches to the normal (i.e., resistive) state.





\section{Results}

In any superconducting device, there exist limits for the supercurrent, $I$, specifically, $I_{c,-}<I<I_{c,+}$. In a non-diode situation, $I_{c,-}=-I_{c,+}$. If a superconducting diode is realized, then $|I_{c,-}|\neq|I_{c,+}|$. Below we show how to construct perfect superconducting diodes (PSD), such that $0<I< I_{c,+}$, i.e., $I_{c,-}=0$. Our model also predicts b-invariant PSD devices in which $I_{c,-}(b)=0$ for a certain finite range of the magnetic field. 

We also propose how to construct devices in which both the lower and the higher limits of the supercurrent are positive, i.e., in which $I_{c,-}>0$. We will refer to these devices as supercurrent range controllers (SRC). They can be used to ensure that the supercurrent stays within certain range and does not cross or approach zero. If the range is narrow, then device can be viewed as a supercurrent stabilizer. 

To obtain such supercurrent rectifiers, the space and time-reversal symmetries must be broken. The space symmetry breaking refers to different critical currents or different critical phases of the wires, or a non-symmetrical distribution of vortices in the cells between the wires. The time-reversal symmetry is broken if a magnetic field is applied or if vortices (i.e., persistent currents in the loops) are present.

\begin{figure}
    \centering
    \includegraphics[width=0.5\linewidth]{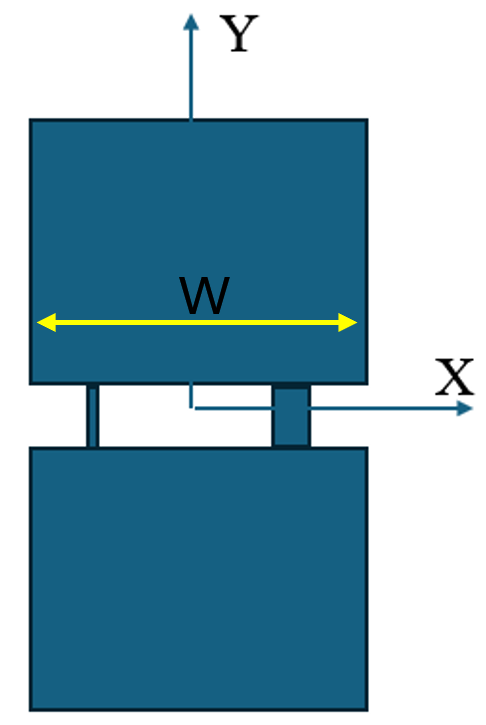}
    \caption{\textbf{Schematic of a two nanowire SQUID (2-SQUID).} It involves two superconducting thin-film electrodes connected by two parallel nanowires. The magnetic field is applied perpendicular to the device, i.e., along the z-axis. One of the nanowires is schematically shown wider to indicate that it has a different $I_c$ and/or a different $\phi_c$ parameters.}
    \label{fig:Sample design}
\end{figure}

\begin{figure}[t]
    \centering
    \includegraphics[width=\linewidth]{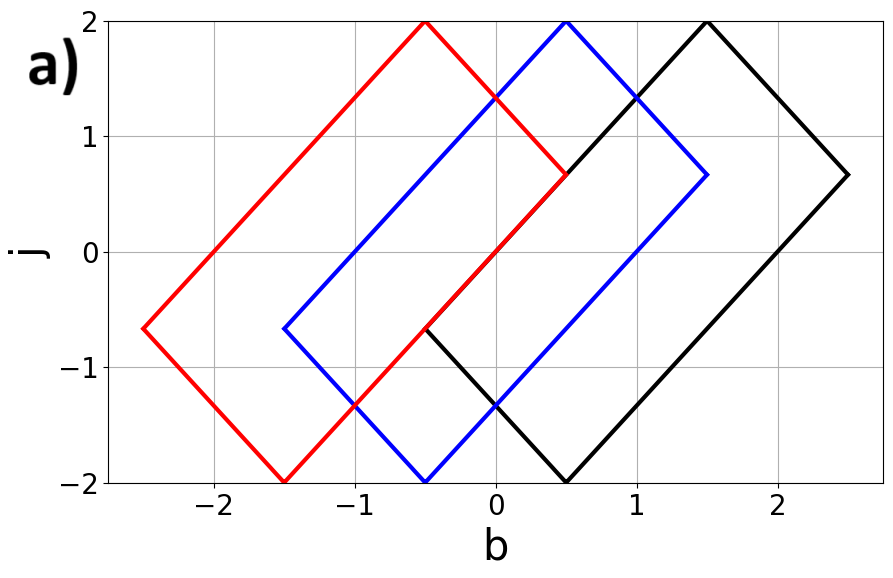}
    \includegraphics[width=\linewidth]{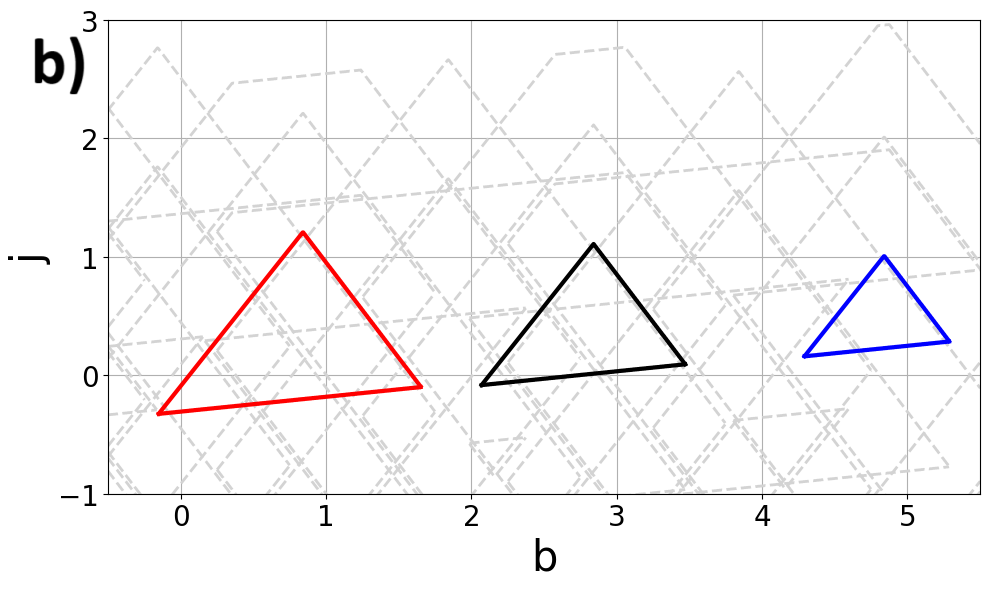}
    \caption{A SRC is demonstrated for VSRs generated by \textbf{(a)} a 2-SQUID with critical phases set to $[\pi, 2\pi]$ and critical currents set to $[2/3, 4/3]$. The plotted vortices are $-1$ (red), $0$ (blue), and $1$ (black). And \textbf{(b)} a disordered 3-SQUID with wires at $[0, 0.45, 1]$, critical phases set to $[6, 7, 5]$, and critical currents set to $[1.01, 1.12, 0.87]$. The vortices from left to right are $[2,-1]$ (red), $[3,0]$ (black), $[4,1]$ (blue). The grey lines indicate any other VSRs corresponding to different vorticity states.}
    \label{fig:abs-psd-example}
\end{figure}

Let us first construct an example of a point-wise PSD. Suppose the SQUID has only two wires, $n=2$, with $v_{1,2}$ vortices trapped between them (see Fig.\ref{fig:Sample design}). Then the MPC equation is $\phi_2=\phi_1 + 2\pi b -2\pi v_{1,2}$. Consider the case $v_{1,2}=0$, so $\phi_2=\phi_1 + 2\pi b$. Zero critical current means that even if the bias current is zero the magnitude of the current in one of the wire is equal to its critical current. Therefore, either $\phi_1=\phi_{c,1}$ or $\phi_1=-\phi_{c,1}$. Suppose $\phi_1=-\phi_{c,1}$. Then $\phi_2=-\phi_{c,1} + 2\pi b$. Under such condition one can increase the bias current and the device will remain superconducting. But a decrease of the bias current is not possible since the first wire would be pushed into the normal state. To construct a PSD, require that the total current is zero, $I=0$, and that the first wire is in the critical state. If $I=0$ then the current in the first wire is equal in magnitude and opposite in sign to the current in the second wire, i.e. $\phi_1I_{c,1}/\phi_{c,1}=-\phi_2I_{c,2}/\phi_{c,2}$. If $\phi_1=-\phi_{c,1}$ then $-\phi_{c,1}I_{c,1}/\phi_{c,1}=-(-\phi_{c,1} + 2\pi b)I_{c,2}/\phi_{c,2}$. So $I_{c,1}=(-\phi_{c,1} + 2\pi b)I_{c,2}/\phi_{c,2}$. For this to be true the right hand side needs to be positive, because $I_{c,1}$ represents the magnitude of the critical current of the first wire. As an example, choose $b = 1$ and $\phi_{c,1}=\pi$. Then $I_{c,1}=\pi I_{c,2}/\phi_{c,2}$. Suppose $\phi_{c,2}=2\pi$. Then $I_{c,1}= I_{c,2}/2$. This argument indicates that a 2-SQUID with critical phases $[\pi, 2\pi]$ and critical currents, for example, $[2/3, 4/3]$ should have a zero critical current at the point $b=1$, i.e., it should act as a perfect superconducting diode. This device will be called point-wise PSD since $I_{c,-}(b)=0$ is only true if $b=1$, or other integers if $v_{1,2}\neq0$ (Fig.\ref{fig:abs-psd-example}a). 

If $v_{1,2}=0$ and if $b>1$, then the allowed range of the supercurrent becomes strictly above zero (Fig.\ref{fig:abs-psd-example}a). For example, if  $b=1.25$ then the allowed range of the supercurrent is $0.3<I<1$. Thus at non-zero magnetic fields our PSD becomes a supercurrent range controller (SRC). The VSR repeats with the period $\Delta b=1$, due to the Little-Parks effect in this 2-SQUID.


Next, we introduce a b-invariant perfect superconducting diode (PSD). Such a device requires a SQUID with more than 2 nanowires $(>2)$. In the b-invariant PSD the critical current of one polarity is independent of the magnetic field (b-invariant), i.e., $I_{c,-}(b)=0$, for $b$ in some finite range. Such a device is realized by a VSR with a flat-top or a flat-bottom, which sits on the $j=0$ axis.

Suppose, as the bias current is increased from zero, the $m$-th nanowire approaches its critical state ($\phi_m=\phi_{m,c}$) earlier than others. Due to the phase correlation imposed by the electrodes, the phase bias on all wires can be expressed through $\phi_m$. The MPC equation is $\phi_i = \phi_m + 2\pi b(x_i - x_m) - 2\pi v_{m,i}$. Assume that all the wires are identical and have the critical current $j_c$ and the critical phase $\phi_c$. The total supercurrent is $j_s = (j_{c}/\phi_c)(\sum_{i=1}^n \phi_i)$.  Therefore 
\begin{equation}\label{eq:total_current}
    j_s = \frac{j_c}{\phi_c}\left(n\phi_m + 2\pi b\sum_{i=1}^n (x_i - x_m)  - 2\pi \sum_i^n v_{i,m}\right)
\end{equation}

For the critical current of one polarity to be independent of the magnetic field, it must be the case that the $b$ term drops out. In other words, $2\pi b \sum_{i=1}^n (x_i - x_m) = 0$ or: 

\begin{equation}\label{eq:nanowire_constraint}
    x_m = \frac{1}{n}\sum_{i=1}^n x_i
\end{equation}
The physical meaning is that the sum of the Meissner phase shifts taken over all wires relative to the $m$-th wire must cancel out. This requirement cannot be satisfied if there are only two wires in SQUID.
Including a nanowire at the average position of the existing nanowires converts the SQUID into a b-invariant PSD. 

Now, assuming that Eq. \eqref{eq:nanowire_constraint} is satisfied. The device is a b-invariant PSD if $I_{c,+}=0$ or if $I_{c,-}=0$. Taking Eq. \ref{eq:total_current} and dropping the $b$ term, we obtain $(j_c/\phi_c)(\pm n\phi_c - 2\pi \sum_{i=1}^n v_{i,m}) = 0$ where $\pm$ indicates the polarity of the critical current that equals zero. Therefore, we obtain a b-invariant PSD if

\begin{equation}\label{eq:PSD}
    \sum_{i=1}^{n}v_{i,m} = \pm \frac{n\phi_c}{2\pi}
\end{equation}

\begin{figure}[t]
    \centering
    \includegraphics[width=\linewidth]{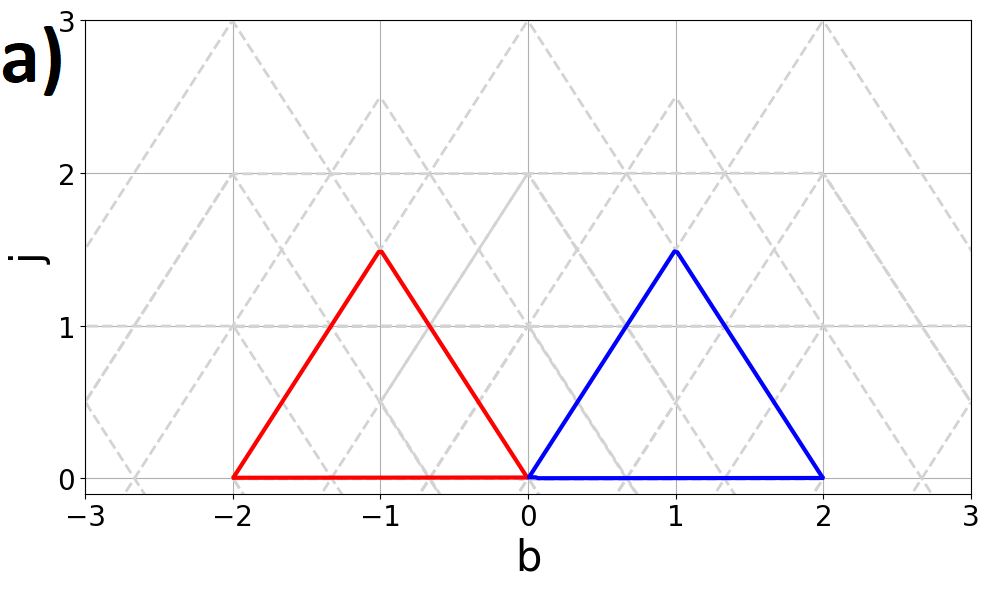}
    \includegraphics[width=\linewidth]{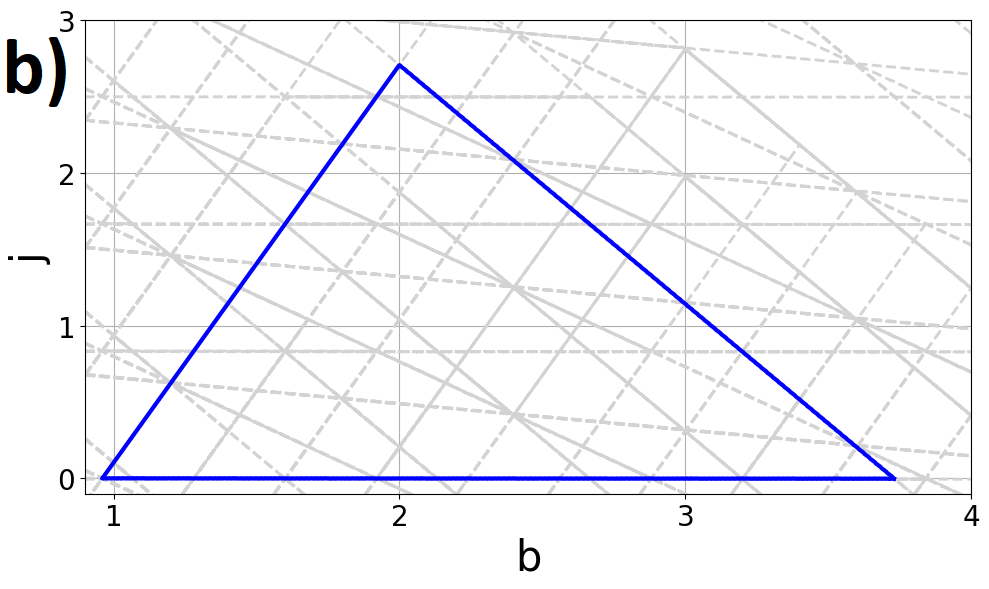}
    \caption{A b-invariant PSD can be constructed by (a) a symmetrical 3-SQUID with $\phi_c=2\pi$ for all wires, setting vorticity states [1,-2] (red) and [2, -1] (blue) and by (b) a non-symmetrical 5-SQUID with wires at $[0,5/8, 2/3, 5/6, 1]$ and critical phases set to $12\pi/5$. The vorticity state is $[3,-1,0,0]$. In both figures, the underlying gray lines indicate other VSRs corresponding to different vorticity states.}
    \label{fig:b-invariant-psd-equidistant}
\end{figure}

\begin{figure}[t]
    \centering
    \includegraphics[width=\linewidth]{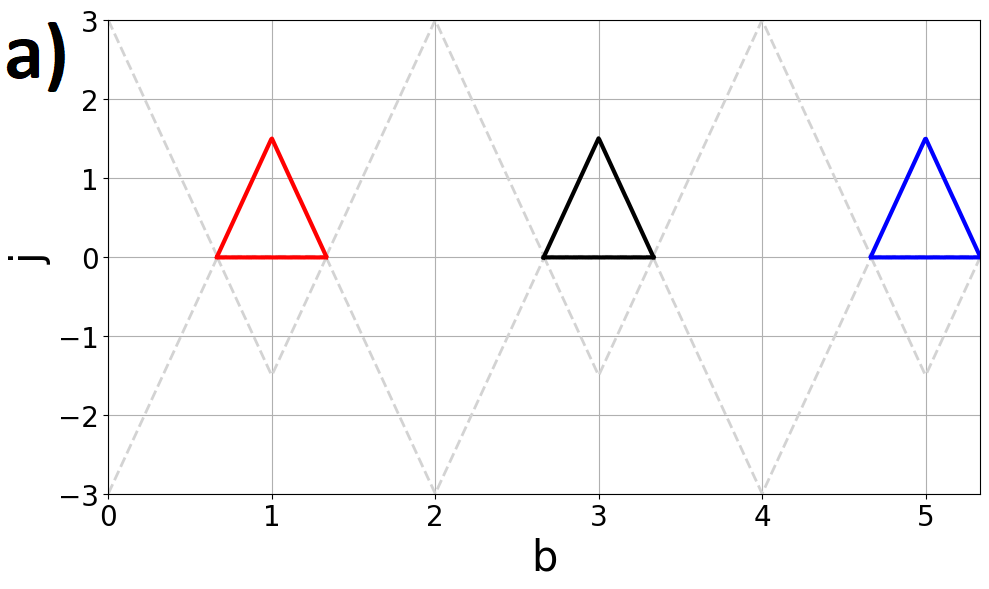}
    \includegraphics[width=\linewidth]{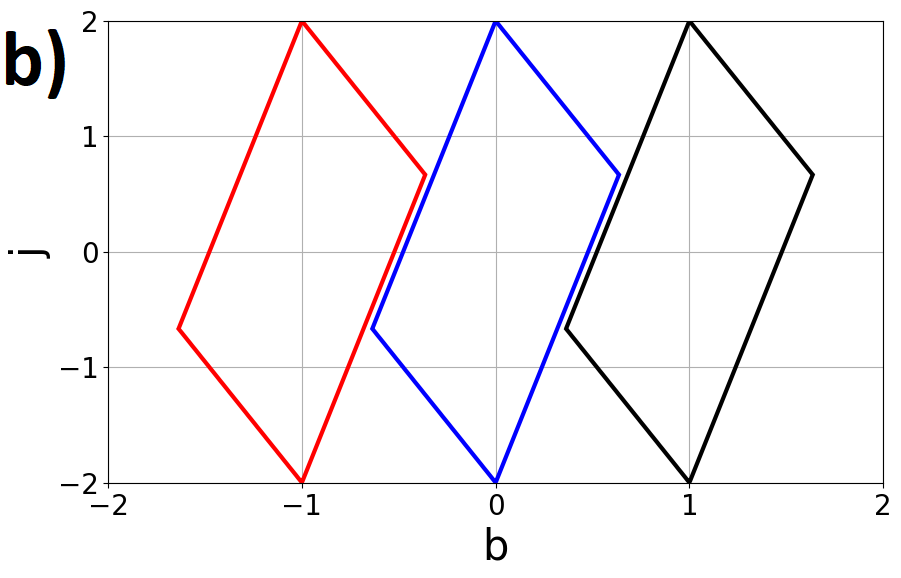}
    \caption{Disjoint VSRs can be leveraged to create an efficient (a) b-invariant PSD from a symmetrical 3-SQUID with $\phi_{c,i}=2\pi/3$ with the vortices [1, 0] (red), [2, 1] (black), and [3,2] (blue) and (b) a SRC from an asymmetrical 2-SQUID with critical phases set to 2 radians and critical currents of $[2/3,4/3]$. The vortices are -1 (red), 0 (blue), 1 (black). In both cases, the critical phase are sufficiently low so that VSRs are disjoint.}
    \label{fig:psd_disjoint}
\end{figure}

 The vorticity state should be programmed such that if $m$ is the switching wire then Eq.\eqref{eq:PSD} applies. Physically speaking, due to existing symmetry breaking, a bias current in one polarity will push the device into the critical regime while a bias current in the opposite polarity will pull the device into the subcritical regime. 

From Eq.\eqref{eq:PSD}, it can be concluded that $\phi_c$ must be an integer multiple of $2\pi/n$, because $\sum_{i=1}^{n}v_{i,m}$ is always an integer.
 As well, Eq.\eqref{eq:PSD} show that the zero vorticity state (zero vortices in all cells) cannot product a b-invariant PSD, despite it being able to produce a SRC. This is because if there are no vortices present in any device cell, then $\sum_{i=1}^{n}v_{i,m} = 0$ which implies $\phi_c = 0$ according to Eq.\eqref{eq:PSD}. Such device is not superconducting at all.
 Therefore, one must always break the geometrical symmetry of the device by introducing vortices to achieve a b-invariant PSD. Additionally, an array of vortices satisfying the constraint $\sum_{i=1}^m v_{i,m} = -\sum_{i=m}^n v_{i,m}$ will also produce an effect equivalent to zero vorticity array. This is the case if, e.g., $n=3$, $m=2$, $v_{1,2}=1$ and $v_{2,3}=1$. Remember that by definition $v_{2,1}=-v_{1,2}$.  This example is a space-symmetrical array because performing a  spatial inversion preserves the equation $\sum_{i=1}^m v_{i,m} + \sum_{i=m}^n v_{i,m} = 0$. Therefore, space symmetry must be broken. This can be achieved be requiring that the cumulative sum of the vorticity numbers is not zero, or, equivalently, the vorticity numbers to the left of the wire $m$ are different from the right side of the device. Therefore, both space and time reversal symmetry must be broken to produce a b-invariant PSD.

Next, we construct an example of a b-invariant PSD in a symmetrical 3-SQUID. Consider a 3-SQUID device with all critical phases set to $2\pi$. According to Eq.\eqref{eq:PSD}, to produce a VSR with the flat and horizontal bottom one needs $\sum_{i=1}^nv_{i,2} = \pm3$ where the 2nd nanowire is the switching wire. We choose to satisfy $\sum_{i=1}^nv_{i,2} = 3$. One such state is $[2,-1]$. The corresponding VSR is shown in Fig.\ref{fig:b-invariant-psd-equidistant}a. It is important to note that the states $[3,0]$, $[1,-2]$, $[0,-3]$, etc. also satisfy $\sum_{i=1}^nv_{i,2} = 3$ while also making the middle nanowire switch first as the bias current of the devices is increased from zero in the positive direction. We show this for the state $v_i = [1,-2]$ with a simple calculation: $\sum_{i=1}^3 v_{i,m} = v_{1,2} + v_{2,2} + v_{3,2} = 1 + 0 + (-1)(-2) = 3$. Recall that $v_{i,m} = -v_{m,i}$ by definition. 
 
 A b-invariant PSD can also be realized from systems with identical superconducting nanowires that are not equidistant. To illustrate this, we did the following: (1) take an array of two rational numbers $[2/3,5/6]$; (2) appended 0 and 1 to this array $[0,2/3,5/6,1]$; (3) computed the average of this array ($5/8$); then (4) placed a 5th nanowire at this array at location $5/8$, so the final locations of the nanowires are $[0, 5/8, 2/3, 5/6, 1]$. This array exemplifies an asymmetrical SQUID. For all wires, choose $\phi_c = 12\pi/5$, so that $ n\phi_c/2\pi=6$ is an integer, where $n=5$. Then, based on Eq.\eqref{eq:PSD}, get $\sum_{i=1}^nv_{i,2} = \pm6$, choose the positive sign. Then, the vorticity state $[3,-1,0,0]$ that satisfies this constraint. Indeed, $\sum {v_{i,m}} = v_{1,2} + v_{2,2} + v_{2,3} + v_{2,4} + v_{2,5} = 3 + 0 + -(-1) + -(-1) + -(-1) = 6$. The 2nd wire is the switching wire (assuming the bias current is ramped up). The VSR corresponding to this state is shown in Fig. \ref{fig:b-invariant-psd-equidistant}b. This is a b-invariant PSD.

The PSDs examples considered above require an exact tuning of the nanowire properties or the magnetic field. However, it turns out that a small disorder in the nanowire location and/or their critical phase converts PSD into a superconducting range controller (SRC). This can be illustrated by introducing disorder into the parameters of the nanowires in a symmetrical 3-SQUID, transforming Fig.\ref{fig:b-invariant-psd-equidistant}a to Fig.\ref{fig:abs-psd-example}b. The VSR triangle is tilted due to the disorder. Recall that the boundaries of the vorticity stability region (VSR) represent the critical currents of the device. Here we observe that the right side of each triangle is lifted. The right triangle (blue) is shifted completely into the positive current domain. Thus the current must be larger than some above-zero threshold in order to flow without any voltage applied. This is the essence of the SRC. Such device permits a supercurrent in a specific  range, $[I_{c,-}>0, I_{c,+}>0]$. This means that the device is \textit{not superconducting} if there is no supercurrent applied to it. This SRC effect is stable against small variations of the device parameters. 

\section{Discussion}

If VSRs overlap, then the device can potentially switch its vorticity state which might be undesirable in some applications. To eliminate the overlap the critical phases has to be as small as possible, although, physically, it is never lower than $\pi/2$. 
Disjoint VSRs have an advantage in that the device is not able to transition to a different vorticity state without going through the normal state. Since the creation of the normal state requires a significant amount of energy, specifically, the condensation energy, such transitions are exponentially suppressed at low temperatures\cite{Sang}. By choosing a low value for $\phi_c$ one can reduces or even eliminate the VSRs overlap because the area of the VSR gets smaller if $\phi_c$ is reduced, as demonstrated below.

If b-invariant PSD is desirable then all the critical phases must be integer multiples of $2\pi/n$, but SRC does not require this condition. For a 3-SQUID b-invariant PSD, if we choose $\phi_c =  2\pi/3$ such that $(3\phi_c)/(2\pi) = 1$, we obtain disjoint VSRs, i.e, for any choice ($j$, $b$) there is only one possible vorticity state (Fig.\ref{fig:psd_disjoint}a). Vorticity states, for which the 2nd wire is the switching wire and which satisfy Eq.\ref{eq:PSD} are $[1,0]$, $[2,1]$, $[3,2]$, etc. For a 2-SQUID SRC, if we choose $\phi_{c,1} = \pi/2$ and $\phi_{c,2} = 2.5$ (rad), the resulting VSRs is also disjoint (Fig.\ref{fig:psd_disjoint}b), i.e., there is always a non-superconducting region between neighbor VSRs.

\begin{figure}
    \centering
    \includegraphics[width=1\linewidth]{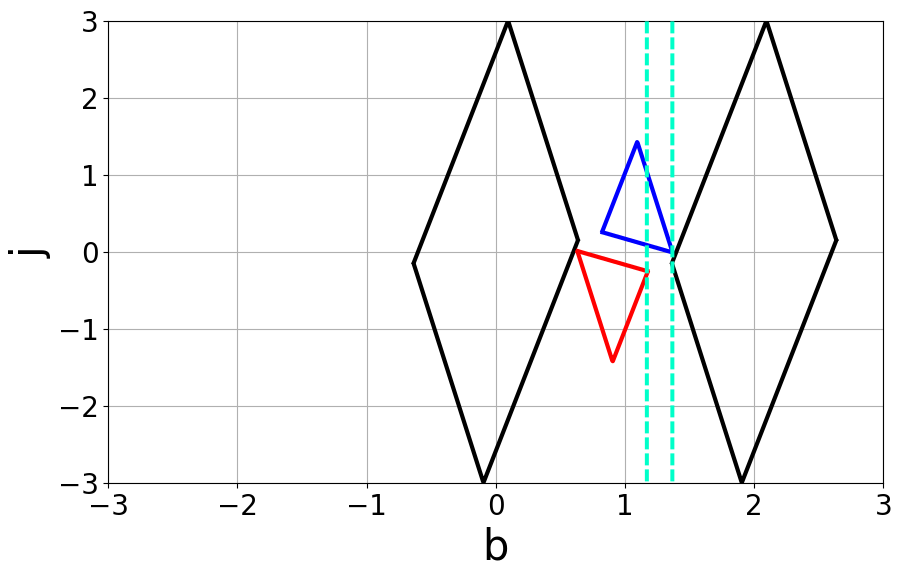}
    \caption{A stable SRC device. Wire locations are [0, 0.5, 1]. Critical phases are [1.7, 2, 2.3] rad. The positive, stable regime is marked by the dashed teal lines.}
    \label{fig:stableSRC}
\end{figure}

An SRC protected against the vorticity switching can also be obtained. In such stable-SRC, if $b$ is between the dashed lines, there is only one possible superconducting vorticity state. See an example in Fig.\ref{fig:stableSRC}.

\section{Conclusion}

A model of superconducting quantum interference device with multiple weak links (nanowires) with a linear current-phase relationship is analyzed. The model predicts that such systems can act as perfect superconducting diodes (PSDs) and/or as superconducting range controllers (SRCs). In PSD, The critical current of one polarity is exactly zero and the critical current of the opposite polarity is above zero ($\eta=1$). In SRC, the supercurrent is bounded between two positive values ($\eta>1$).  A point-wise PSD device may be realized with 2-SQUIDs whereas a b-invariant PSD device is realized with SQUIDs with $>2$ wires. In contrast, a SRC device may be realized with 2-SQUID. A small difference between the nanowires does not eliminate the strong diode effect reported here, but it changes PSD into SRC.

\begin{acknowledgments}
The work was supported in part by the NSF DMR-2104757
and by the NSF OMA 2016136 Quantum Leap Institute for
Hybrid Quantum Architectures and Networks (HQAN). We also like to thank the University of Illinois Materials Research Laboratory for providing us with capabilities of sample fabrication and imaging.

\end{acknowledgments}

\section*{References}
\bibliography{references}
\end{document}